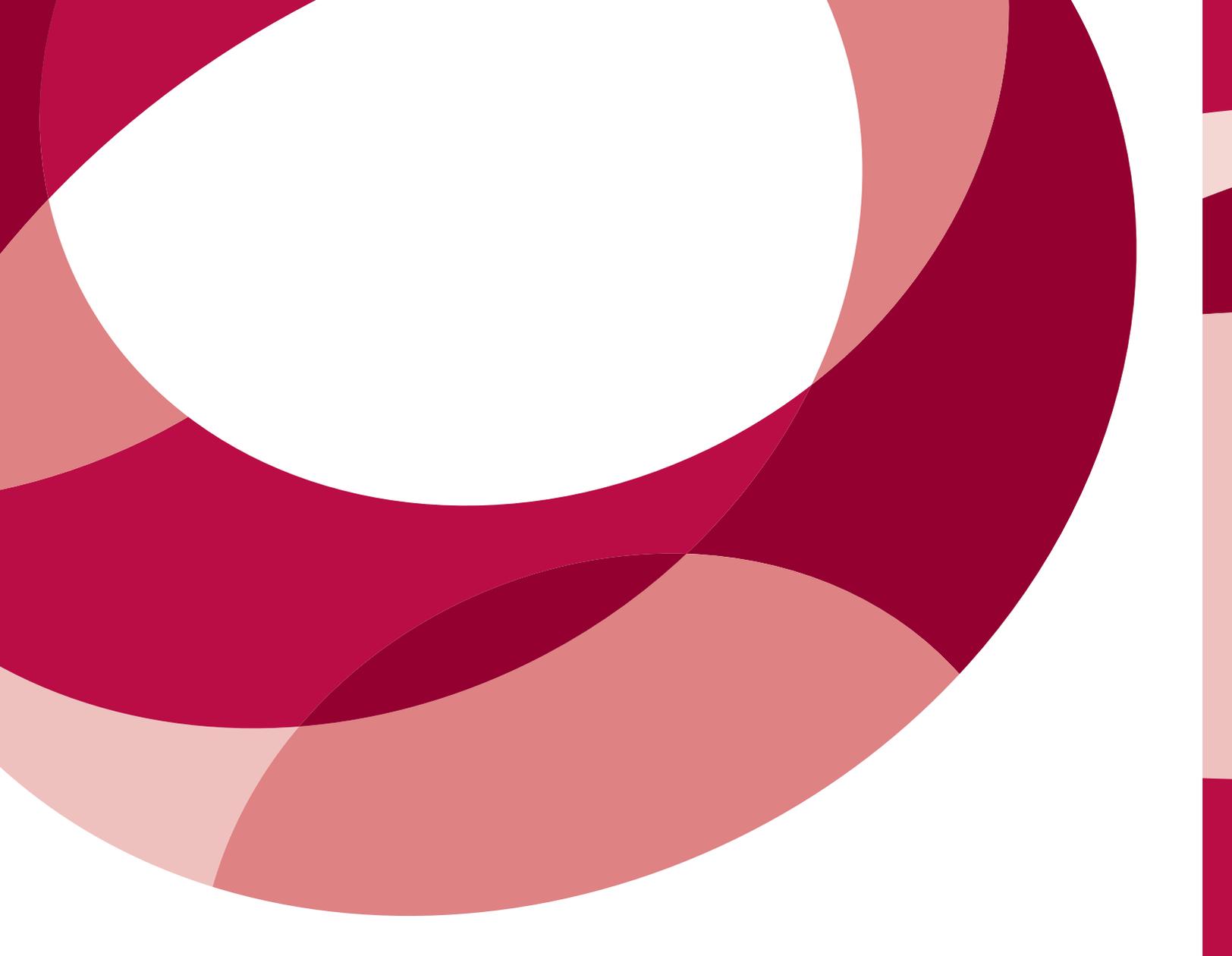

# The Future of Research on Social Technologies: CCC Workshop Visioning Report

April 2024

The workshop was supported by the Computing Community Consortium through the National Science Foundation under Grants No. 1734706 and 2300842, with additional support from the John S. and James L. Knight Foundation. Any opinions, findings, and conclusions or recommendations expressed in this material are those of the authors and do not necessarily reflect the views of the National Science Foundation.

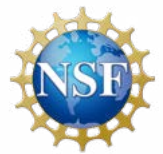

# The Future of Research on Social Technologies: CCC Workshop Visioning Report

# April 2024

**Workshop Organizers**[1]

Motahhare Eslami, Carnegie Mellon University
Eric Gilbert, University of Michigan
Sarita Schoenebeck, University of Michigan

**Contributors**

Eric P. S. Baumer, Lehigh University
Eshwar Chandrasekharan, University of Illinois at Urbana-Champaign
Michelle De Mooy, Georgetown University
Karrie Karahalios, University of Illinois at Urbana-Champaign
David Karger, Massachusetts Institute of Technology
Tressie McMillan Cottom, University of North Carolina at Chapel Hill
Andrés Monroy-Hernández, Princeton University
Loren Terveen, University of Minnesota
John Wihbey, Northeastern University

---

[1] The report was written by the authors listed in Organizers and Contributors. Generative AI systems were not used.

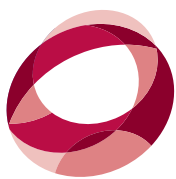

# Executive Summary

Social technologies are the systems, interfaces, features, infrastructures, and architectures that allow people to interact with each other online. These technologies dramatically shape the fabric of our everyday lives, from the information we consume to the people we interact with to the foundations of our culture and politics. While the benefits of social technologies are well-documented, the harms, too, have cast a long shadow. To address widespread problems like harassment, disinformation, information access, and mental health concerns, we need to rethink the foundations of how social technologies are designed, sustained, and governed.

This report is based on discussions at the Computing Community Consortium (CCC) Workshop "The Future of Research on Social Technologies" that was held November 2-3, 2023 in Washington, DC. The visioning workshop came together to focus on two questions: What should we know about social technologies, and what is needed to get there? The workshop brought together over 50 information and computer scientists, social scientists, communication and journalism scholars, and policy experts (see Appendix A). We used a discussion format, with one day of guiding topics and a second day using an "unconference" model where participants created discussion topics. The interdisciplinary group of attendees discussed gaps in existing scholarship and the methods, resources, access, and collective effort needed to address those gaps. We also discussed approaches for translating scholarship for various audiences—including citizens, funders, educators, industry professionals, and policymakers. As part of this, we worked with a graphic recorder who produced artwork in real-time that highlights many key points of workshop discussions (see Appendix B).

This report presents a synthesis of major themes during our discussions. The themes presented are not a summary of what we know already; they are an exploration of what we do not know enough about, and what we should spend more effort and investment on in the coming years.

◗ The first theme focuses on rethinking the design of social platforms and infrastructures, and explores alternative models involving decentralization, distributed moderation, governance, and monitoring and measurement of a platform/community's "health."

◗ The second theme reflects on opportunities at the intersection of social technologies and democracy—including community, corporate, and regulatory layers of governance, and how technology and regulation might protect and defend democratic principles across these layers.

◗ The third theme focuses on new challenges for social technologies and AI, including authenticity, ranking, and social training data.

◗ The fourth theme focuses on current challenges for research access and practice, which has changed significantly in recent years.

◗ The final theme addresses impact and translation. It calls attention to the importance of training computing researchers to translate work for broader audiences and incentives to do translation work. Similar to other societally important areas of computing, pursuing these opportunities requires thinking beyond individual papers; we should explore cross-institutional collaborative models, fund alternative approaches to platforms, and shift academic incentives to high-impact work.

## Summary of Opportunities (by theme)

### 1. Rethinking Social Platforms

1.1. Researchers can help inform the governance of social platforms in ways that are social, technical, and legal.

1.2. Researchers can significantly inform the trajectory of emerging decentralized social platforms, over the near- and long-term.

1.3. Researchers can build methods—both quantitative and qualitative—to measure and assess the "health" of social platforms and online communities.

### 2. Social Technologies and Democracy

2.1. Researchers can significantly shape the design and study of social technologies to encourage democratic principles and protect against threats.

2.2. The research community should invest in mutual support and stand up for threatened researchers, and advocate for strong protection for societally important, yet threatened, research at their home institutions.

### 3. Social Technologies and AI

3.1. Researchers should investigate and design mechanisms to use AI for prosocial ends in online communities.

3.2. Researchers have an opportunity to build and test new designs for social platforms that recognize the presence of AI-generated content.

3.3. Researchers can help safeguard certain social data from appropriation by AI systems, and AI systems from certain social data.

### 4. Research Access and Practice

4.1. Researchers should study both big platforms and small, niche communities with respect to data access opportunities.

4.2. Researchers should explore opportunities to develop shared infrastructure and practices for data collection and sharing, and experimental work.

### 5. Impact Beyond Academia

5.1. The research community can increase the potential for impact by shifting incentive structures to recognize diverse kinds of work and contributions, particularly high-impact, large-scale, and/or community-centered work.

5.2. Researchers can explore new models of research that move beyond individual researchers towards collective efforts.

5.3. Researchers have an opportunity to impact policy related to social technologies.

## 1. Rethinking Social Platforms

A great deal of discussion during the workshop centered on new visions for social platforms, their infrastructures, and how to govern them. Specifically, many workshop participants see an opportunity in the current chaos around social media: opportunity to envision, build, and study new models and architectures for social platforms. New models could use decentralized or public architectures, or ones that have explicitly research-informed approaches to governance from the start. Multiple participants raised the questions: How do we know if we're heading in the right direction? How do we know if a social platform is getting better, and for whom?

### Governing Online Behavior

Platform governance can happen at three layers: community, corporate, and regulatory. Community governance puts the power in the hands of the people in the community. This allows them to make decisions about what kinds of values, norms, and content they want to encourage and allow. However, not all community members want to have this much responsibility. For example, Wikipedia works because it relies on the labor of its community members, but it's not clear how reproducible and scalable Wikipedia's model is for other platforms. Additionally, communities may not have enough power to meaningfully make or enforce their own rules, especially if they are operating under the umbrella of corporate policies. Even the very meaning of "governance" can be conceptually slippery in these contexts: for example, who governs email?

The corporate layer is the status quo currently—most platforms are run by corporations and they develop policies about what users can do and not do. These policies are increasingly scrutinized by lawmakers, as concerns about the harms of social media continue to grow. There could be regulatory bodies that audit corporate governance practices and consumer wellbeing. However, it may be hard to push for substantial changes, given the politicized nature of technology policy and that users often like their current experiences.

Still, there is a lot of potential to imagine new forms of governance at the community, corporate, and regulatory levels. This could mean many different things: governing platforms with existing and emerging laws (e.g., the Digital Services Act in the EU), building toolkits to allow users to self-govern either existing or new architectures, or monitoring platforms sufficiently so that governments can use and enact laws to govern them effectively.

**1.1 High-level opportunity: Researchers can help inform the governance of social platforms in ways that are social, technical, and legal.**

*1.1.a Concrete opportunity: Sociotechnical research has an opportunity to "plug into" emerging laws such as the DSA in the EU (e.g., vetted researcher access to data from very large platforms). Given the structure of these and other emerging laws, new partnerships with the EU will need to be built.*

*1.1.b Concrete opportunity: Researchers have the opportunity to help define and build new governance structures for existing and emerging platforms—ranging from collective decision-making around algorithmic systems that mediate interactions to community moderation.*

*1.1.c Concrete opportunity: Monitoring social platforms continues to be a challenge, but one that is necessary for science-informed tech policy. Building (and defending) toolkits to continually audit social platforms remains a significant opportunity for computing research.*

### Decentralization

There are emerging social platforms that embrace decentralization as an architectural and philosophical value. Unlike corporate-controlled and centralized systems such as Facebook or Instagram or X, decentralized architectures are often either physically or administratively decentralized: they may permit data to be stored all over the internet and routed through standardized protocols, or allow distributed groups of people to govern themselves with distinct rules.[2] Current examples include: Bluesky, which uses the AT protocol; Mastodon, which uses the ActivityPub W3C standard; and, Farcaster, which uses the blockchain-based Farcaster protocol. It is likely that others will emerge over the next few years.

The research community is well-positioned to understand, support, and build for these emerging platforms. In addition to many open questions surrounding how to best build systems like these, decentralized architectures are fundamentally more "open" to researchers, compared to corporate platforms which have become more closed in recent years. However, decentralization creates new challenges and opportunities. Economic forces often drive forms of "de facto centralization" in decentralized systems (e.g., economies of scale, spam, etc.). Moreover, in centralized architectures, those who control the platform can enforce moderation and governance decisions on those who use the platform. In decentralized architectures, different components of the system may be controlled by different entities who make their own enforcement choices. This can make it far more difficult to prevent the dissemination of content (at scale), which has consequences both good (less censorship) and bad (more harassment and misinformation). Additionally, in a decentralized environment, we may need to worry not only about individual bad-actor *users* but about bad-actor servers that aim to subvert the entire ecosystem by hiding, revealing, misdirecting, or fabricating communications. Finally, decentralization makes it far more difficult to deliver a global view of the entire ecosystem.

---

[2] Reddit might be considered to be (somewhat) administratively decentralized, for example.

**1.2 High-level opportunity: Researchers can significantly inform the trajectory of emerging decentralized social platforms, over the near- and long-term.**

*1.2.a Concrete opportunity: Researchers can significantly inform the architecture and understanding of distributed moderation, governance, and curation—where the meaning of those terms may be less clear.*

*1.2.b Concrete opportunity: Researchers can inform and combat the under-explored security risks that arise in distributed social platforms.*

*1.2.c Concrete opportunity: Researchers can help to design scalable and sustainable economic models for decentralized architectures.*

### Measurement and Evaluation

What does it mean to make a social platform *better*? Multiple participants noted that we lack a scientific approach for measuring and assessing whether an online community is "healthy," and whether a design change moves it in a better direction. For example, does changing a feed ranking algorithm in a specific way make a community more resilient to misinformation? Does a change to the presentation of identity in an online community lead to more prosocial behavior? Researchers and designers employed by platforms will often assess changes like these via massive A/B experiments—but usually with limited outcome metrics, such as "seconds on platform." The research community is well-positioned to think about this problem much more holistically: this approach could help build science, while also building tools for industry designers and scientists, and perhaps for future oversight bodies. The research community must also recognize that metrics are a reflection of values.

**1.3 High-level opportunity: Researchers can build methods—both quantitative and qualitative—to measure and assess the "health" of social platforms and online communities.**

*1.3.a Concrete opportunity: Researchers can explore new approaches to algorithmically measure and model how healthy an online community is, and how it may change over time; a particular challenge will be whether important nuances of online interactions can be captured algorithmically.*

*1.3.b Concrete opportunity: Researchers can explore new approaches to qualitatively assess how healthy an online community is; a particular challenge will be speed and scalability of such a method.*

*1.3.c Concrete opportunity: Researchers can prioritize the experiences of marginalized and underrepresented users in its assessments of online community health.*

*1.3.d Concrete opportunity: Researchers should explore how to balance societal concerns in addition to user concerns, including when societal concerns may be at odds with users' expressed preferences at a given moment.*

## 2. Social Technologies and Democracy

Early scholarship at the intersection of social technologies and democracy focused on the power of digital affordances to facilitate collective action, around everything from local government to civic programs, and from civil society protests to creating structures for providing input to government. Yet there has been a decisive turn in this subfield toward a more critical stance, as harms have become more apparent across many domains, and the use of technological tools for anti-social, anti-democratic ends has become more visible. Further, the role of technological design, whether at the interface or algorithmic level, has come under much more scrutiny in terms of its impact on democracy-related outcomes.

While election integrity remains an area of focus in terms of evaluating the impact of social technologies on democracy, there remain broader questions about how technologies allow participation in civic life generally and facilitate representation in the public sphere. Scholars have increasingly focused on how social technologies serve to empower or disempower voices and how they may exacerbate marginalization and create unsafe environments. The study of online disinformation and hate speech often speaks to core democratic issues of participation in the online public sphere and the ability of societies to make informed decisions.

Around the world, elections are experienced and shaped by online information ecosystems. Across countries and languages, there are concerns about election interference and low information quality. Concerns about the ability to organize around anti-democratic and authoritarian goals are also salient. Novel challenges exist around how to handle cross-platform campaigns, as well as how alternative, non-mainstream platforms with little governance in place might be studied, understood, and handled from the perspective of systemic risk to democratic institutions. Relatively little research has been done, and data collected, relating to niche, alternative social platforms that host many of the more polarizing or extreme groups. The rise of encrypted platforms—while serving important purposes—also makes such research all-the-more difficult.

It also remains unclear if scholars are studying high-stakes concerns like polarization and extremism in a comprehensive way, and how it may be measured more generally with regard to social technologies. Holding global platforms to account for their lack of effective content moderation has become a key issue for scholars studying human rights and the protection and safety of minority groups around the world. Finally, much of the research related to elections and democracy online is reactive, studying what happened after the fact. More research is needed on how to proactively detect potential threats to democracy and shift the designs, norms, and policies of a platform to preserve democratic principles.

**2.1 High-level opportunity: Researchers can significantly shape the design and study of social technologies to encourage democratic principles and protect against threats.**

*2.1.a Concrete opportunity: Researchers should study the treatment of marginalized voices globally online, especially during high-stakes periods like elections. New technical and empirical work may offer new ways to do this.*

*2.1.b Concrete opportunity: Researchers should devote effort to study alternative platforms that may be "under the radar," but may have a substantial impact on democratic processes and principles.*

### Engagement and Risks

At the same time, we recognize that doing research in this domain carries risk. Members of our research community have been harassed, threatened, and had their work mischaracterized in partisan media and governmental reports. Academics and industry practitioners have been (and continue to be) investigated, subpoenaed, and sued. They have also been made visible (and put into a negative spotlight) by high-profile industry actors, partisan media outlets, and government officials. Some institutions have been able to marshall resources to protect their employees, but not all institutions have the will or capacity to do so. Moreover, some researchers work outside traditional institutions, or may not have sufficient standing at their institution to receive such protection (e.g., students vs. tenured faculty at elite R1 institutions).

While these risks are significant, the work is important. There may be opportunities for the community to form networks and communities of mutual support, and to advocate within and across their institutions for strong protection of academic freedom as declared by the AAUP's "1940 Statement of Principles on Academic Freedom and Tenure." Finally, senior faculty should stand up for and vocally support researchers who undertake such important work.

**2.2 High-level opportunity: The research community should invest in mutual support and stand up for threatened researchers, and advocate for strong protection for societally important, yet threatened, research at their home institutions.**

## 3. Social Technologies and AI

The workshop extensively discussed Artificial Intelligence (AI), particularly the rise of generative AI technologies, and the impact of AI on social technologies—and vice versa. This included several opportunities for the social computing community, given the unique challenges of human-AI design. Discussions often returned to the future of social platforms in the presence of AI, particularly generative AI and its potential applications. Participants talked about the relevance of the social computing field with the recent advancements in the AI field, and what it means to be social in the AI era. At the same time, there are also many ethical regulatory concerns that have yet to be addressed.

### Increasing Resources for Moderation

Although questions about the harms of algorithmic curation remain to be answered, there is potential to make algorithmic curation beneficial to tackle different forms of online antisocial behavior, including online misinformation (e.g., rabbit holes of misinformation via algorithmic recommendations), echo chambers, hate speech, and harassment. Participants reflected on how AI can be developed to be a defender of the information environment. In addition to more audits of current algorithmic approaches—like algorithmic curation within feeds (e.g., ranking, filtering, recommendations)—further research is needed to explore how existing approaches can be improved and how new technologies can be developed to leverage algorithmic socio-technical affordances.

**3.1 High-level opportunity: Researchers should investigate and design mechanisms to use AI for prosocial ends in online communities.**

### Authenticity and Veracity

The workshop discussed distinguishing AI-generated content from human-generated content, given generative AI's rise to prominence. The emergence of generative AI-powered chatbots makes it challenging to distinguish between content generated by AI from humans. To date, this has largely been viewed as an *AI problem*. Many participants argued that there is a large opportunity for social computing to help address it as well. Most of the popular generative AI systems (e.g., ChatGPT, Gemini, Claude) are trained using multiple internet sources including data generated by users of social technologies. This can result in authenticity challenges, and lack of trust in online interactions.

Furthermore, there is concern about the raw scale of AI-generated content. Participants argued that with the pace of AI-generated content, social spaces could potentially be filled with "garbage." Will AI inevitably swamp human-generated content? What will happen to internet communities if and when they face a deluge of AI-generated content? Will community members, moderators, and admins be able to tell the difference? Will it matter if they cannot? These and similar questions are central in the coming years, as new AI technologies begin influencing the structure and dynamics of online communities and social platforms.

**3.2 High-level opportunity: Researchers have an opportunity to build and test new designs for social platforms that recognize the presence of AI-generated content.**

*3.2.a Concrete opportunity: Researchers can help design and implement robust verification systems that can authenticate the entity on the other end—AI or human. This could involve mechanisms including digital signatures, verification badges, collective fact checking, and trust mechanisms.*

*3.2.b Concrete opportunity: Researchers can help develop standards for indicating the provenance of information online, including tracking where and how it was generated.*

### Social Data and AI

AI systems are trained extensively on social data from the internet. That is, AI systems have been trained on the text, images, video, and other media produced by people on the internet over the last 30 or so years. The ethics and legality of this is highly debatable, and at the time of this writing, is the focus of multiple high-profile court cases in the U.S. and elsewhere. Broadly speaking, participants argued that there may be opportunities for the social computing community to contribute in two ways. First, researchers may be able to build systems that can safeguard certain types of data and media from being appropriated by AI systems (e.g., artwork by independent artists). Second, it is well-established that not all online data are worth emulating. How do we teach this to AI systems, systems that presently have few ways of distinguishing between what should be learned and what should not?

**3.3 High-level opportunity: Researchers can help safeguard certain social data from appropriation by AI systems, and AI systems from certain social data.**

*3.3.a Concrete opportunity: Researchers can build systems that safeguard certain content from being used in training in novel ways. This may include building databases to which creators can deposit their work, and new technical means of watermarking content.*

*3.3.b Concrete opportunity: Research can help to build new algorithmic approaches for teaching AI systems which online social data they should emulate, and which they should not.*

## 4. Research Access and Practice

Access to platforms and the people who use them is essential for our work. Many of those issues have been discussed in the preceding sections. In addition, many participants stressed the importance of creating and maintaining new social computing infrastructures—for research and practice. Yet, doing so requires a critical mass of subscribers and financial support for development and maintenance. For example, the field cannot do research without users, and users won't participate in a platform that is going to sunset when the research ends. Such infrastructures need to exist for a minimum number of years to attract and sustain a network of people, and they take time and luck to develop. While exploring this option, discussions revealed new international opportunities for data access, standardization, and industry engagement.

### Data Access

Data access around social platforms has often centered around researchers' access to corporate data from select, large organizations. Challenges include limited access to data (e.g., lack of APIs to collect data, chilling effects around non-API data collection), data access only given to a small group of researchers, quality of data collection, lack of regulation, and more. Companies publish some of their policies and decisions, but only selectively and often on their own terms. While corporate data is important for studying the effects and impacts of social technologies, we should not be so reliant on only that approach.

**4.1 High-level opportunity: Researchers should study both big platforms and small, niche communities with respect to data access opportunities.**

*4.1.a Concrete opportunity: Researchers need to explore smaller, diverse communities, outside just the big corporate platforms, and expand the types of platforms and technologies they focus on obtaining data from and with.*

*4.1.b Concrete opportunity: Researchers should explore international data sources and partnerships that are outside North America and Europe.*

**Standardization**

Standardization also emerged as a theme around infrastructures, data, and regulation. There is a lot of "reinventing the wheel" in the community: researchers often build bespoke architectures for a single or small number of research projects that do not get shared back with the community.

**4.2 High-level opportunity: Researchers should explore opportunities to develop shared infrastructure and practices for data collection and sharing, and experimental work.**

*4.2.a Concrete opportunity: Researchers should explore data archives for storing social platform research data according to standardized schemas—that perhaps becomes a de-facto requirement for publishing, as it is in fields like biology.*

*4.2.b Concrete opportunity: Researchers should develop standards in social computing infrastructures, data sharing, and audits to encourage and facilitate collective development and assessment.*

## 5. Impact Beyond Academia

Our discussions routinely returned to the question of impact. While academic research is already known to be siloed, tucked away in scholarly journals, participants felt a particular urgency and responsibility to translate research on social technologies into public impact.

**Incentives in Academia**

Academic researchers are subject to a variety of constraints that challenge their ability to conduct research that has an impact beyond academia. Many of these constraints arise from the incentive structures in academia, which often prioritize scholarly publications over most other activities. Research is typically codified as the publication of research papers in scholarly journals and conference proceedings. We have an opportunity to think differently, and think bigger. We might incentivize different kinds of research, such as community-based research which takes a long time and may yield fewer publications, but which can have meaningful impact within communities. We could incentivize translational work, that involves translating each scholarly publication into a format that is accessible and useful to stakeholders outside of academia. Doing this kind of work takes time and would require shifting incentives around scholarly publications, such as only evaluating the few more important papers during merit and promotion processes.[3] In light of the topic areas we have discussed above (e.g., building entirely new systems for social platforms, deep engagement with how social technologies affect democracy), the community needs to appropriately weight big contributions over smaller, more incremental projects and papers.

**5.1 High-level opportunity: The research community can increase the potential for impact by shifting incentive structures to recognize diverse kinds of work and contributions, particularly high-impact, large-scale, and/or community-centered work.**

*5.1.a Concrete opportunity: The research community can incentivize impact by focusing on the quality and impact of a few publications, rather than focusing on quantity of publications, during evaluation and promotion processes. This is a norm that could emerge in letter writing for promotion and tenure, for example.*

*5.1.b Concrete opportunity: The research community can incentivize different types of scholarly engagement, such as community-based research, by recognizing the time and resources it requires.[4]*

*5.1.c Concrete opportunity: Researchers can develop field- and discipline-specific guidance on measuring, tracking, and evaluating extra-institutional impact.*

**Rethinking Knowledge Production**

The workshop discussed approaches to conducting research that move away from individualistic models towards more collective, organizational structures. This could involve research groups collaborating in more structured aways across institutions, connecting researchers with community organizations, and connecting researchers and community members. There are also initiatives that provide expertise to researchers, such as statistical consulting, legal consulting, or computational infrastructure.

---

[3] https://cra.org/resources/best-practice-memos/incentivizing-quality-and-impact-evaluating-scholarship-in-hiring-tenure-and-promotion/

[4] http://cra.org/ccc/wp-content/uploads/sites/2/2024/03/CDARTS-Workshop-Report_Final.pdf

At the same time, unorthodox configurations can produce tensions. For example, community organizations have different priorities than researchers and it can be hard to align them in a collaboration. There can also be power dynamics that assume academic researchers have the necessary expertise or knowledge, when in fact it may be the other way around. Finally, academics may take steps to make their work more public and transparent, such as putting code and results on GitHub, but this does not necessarily mean other communities can make effective use of it.

**5.2 High-level opportunity: Researchers can explore new models of research that move beyond individual researchers towards collective efforts.**

*5.2.a Concrete opportunity: Researchers can develop principles for doing cross-organizational work (e.g., various stakeholders) in ways that are mutually beneficial rather than extractive.*

### Policy Relevance

The intersections between social technologies and policy have become much stronger and more salient in recent years, and many participants were interested in those intersections. With catalyzing events like Cambridge Analytica, global presidential elections, or Elon Musk's takeover of Twitter/X, many academics now see policy considerations as within the purview of their work. What some participants are less clear on is how, and to what extent, to engage with policy work. This includes questions about how to translate research results to policy recommendations, and how to actually do the work of communicating with policymakers and the broader public. We discussed mechanisms for doing policy work, including writing op-eds or blog posts, getting in front of reporters, or talking with staffers and politicians. We note that this is distinct and different from lobbying activities.

Policy work can be more effective when done collectively, by working with other colleagues at a researcher's own institution or across institutions or by working with civil society organizations. Academics tend to fixate on details, but this can be less effective when translating to policy. Participants who came from the policy world recommended focusing on clear messages rather than being too "in the weeds" of academic content.

**5.3 High-level opportunity: Researchers have an opportunity to impact policy related to social technologies.**

*5.3.a Concrete opportunity: Researchers can learn how to translate research to inform policy decisions, how to communicate with policymakers, and how to build partnerships to increase the likelihood of impact. The field should explore mechanisms by which we can accelerate that impact, such as training and summer schools.*

## 6. How to Cite this Report

Eslami, M., Gilbert, E., Schoenebeck, S., Baumer, E. P. S., Chandrasekharan, E., De Mooy, M., Karahalios, K., Karger, D., Cottom, T. M., Monroy-Hernández, A., Terveen, L., & Whibhey, J. (2024). (rep.). *The Future of Research on Social Technologies CCC Workshop Visioning Report*. Computing Community Consortium.

## Appendix A: Workshop Participants

| First Name | Last Name | Company Name |
| --- | --- | --- |
| Kendra | Albert | Cyberlaw Clinic, Harvard Law |
| Eric | Baumer | Lehigh University |
| Michael | Bernstein | Stanford University |
| Eshwar | Chandrasekharan | University of Illinois, Urbana- |
| A.J. | D'Amico | Knight Foundation |
| Elora | Daniels | CRA |
| Michelle | De Mooy | Georgetown University |
| Judith | Donath | Berkman-Klein Center at |
| Joan | Donovan | Boston University |
| Will | Duffield | Cato Institute |
| Brent M. | Eastwood | R Street Institute |
| Motahhare | Eslami | Carnegie Mellon University |
| Casey | Fiesler | University of Colorado |
| Susan | Fussell | Cornell Univ |
| Greg | Gersch | Greg Gersch LLC |
| Eric | Gilbert | University of Michigan |
| Cat | Gill | CRA |
| Cami | Goray | University of Michigan |
| Kishonna | Gray | University of Kentucky |
| Haley | Griffin | CCC |
| Peter | Harsha | Computing Research |
| Anna | Harvey | Social Science Research Council |
| Benjamin Mako | Hill | University of Washington |
| David | Jensen | University of Massachusetts Amherst |
| Karrie | Karahalios | University of Illinois, Urbana-Champaign |
| David | Karger | MIT |
| Seyun | Kim | Carnegie Mellon University |
| Laura | Kurek | University of Michigan |
| Anna | Lenhart | Institute for Data Democracy and Politics |
| Tom | Martin | National Science Foundation |
| Winter | Mason | Meta, Inc. |
| Tressie | McMillan Cottom | UNC Chapel Hill; New York Times |
| Danaë | Metaxa | University of Pennsylvania |
| Tanu | Mitra | University of Washington |
| Andrés | Monroy-Hernández | Princeton University |
| Mor | Naaman | Cornell Tech |
| David | Nemer | University of Virginia |
| Michael | Pozmantier | National Science Foundation |
| Yoel | Roth | University of Pennsylvania |
| Zeve | Sanderson | NYU Center for Social Media & Politics |
| John | Sands | John S. and James L. Knight Foundation |
| Ann | Schwartz | CRA |
| Sarita | Schoenebeck | University of Michigan |
| Katie | Siek | Indiana University |
| Loren | Terveen | The University of Minnesota |
| John | Wihbey | Northeastern University |
| Pam | Wisniewski | Vanderbilt University |
| Ashley | Zohn | Knight Foundation |
| Ethan | Zuckerman | UMass Amherst |

## Appendix B: Graphic Recordings of Key Points of Workshop Discussions

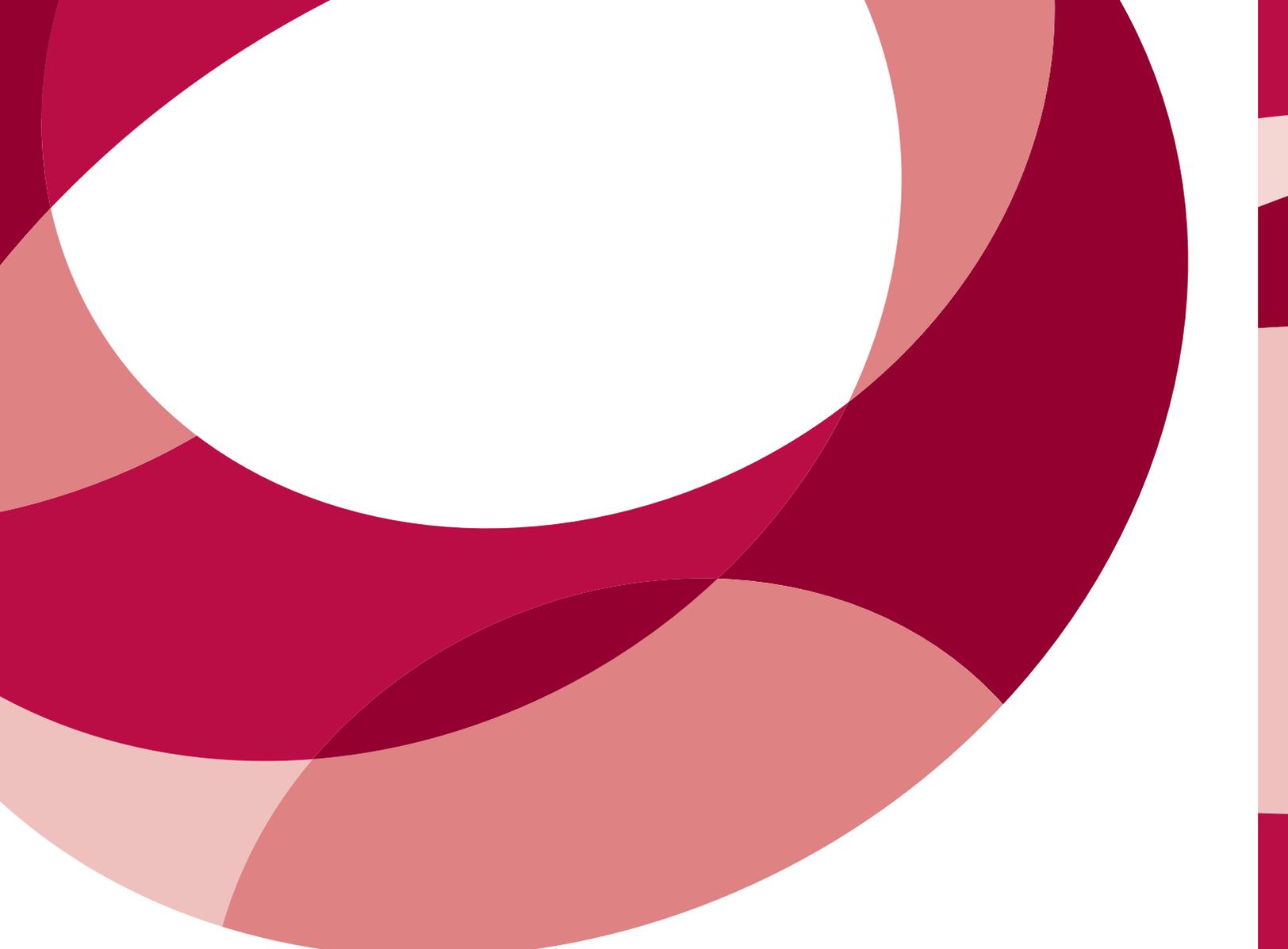

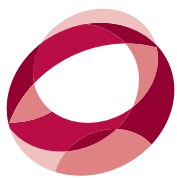
CCC
Computing Community Consortium
Catalyst

1828 L Street, NW, Suite 800
Washington, DC 20036
P: 202 234 2111  F: 202 667 1066
www.cra.org/ccc  cccinfo@cra.org